\documentclass[11pt,letterpaper]{article}
\usepackage[T1]{fontenc}
\usepackage[latin9]{inputenc}
\usepackage{textcomp}
\usepackage{amsmath}
\usepackage{amssymb}
\usepackage{esint}
\usepackage{capt-of}
\usepackage{physics}

\usepackage{tikz, pgfplots}
\usepackage{tikz-3dplot}
\usetikzlibrary{arrows, shapes.misc}
\usetikzlibrary{decorations.markings}
\usetikzlibrary{decorations.pathmorphing}
\usetikzlibrary{backgrounds,automata}
\pgfplotsset{compat=newest}

\makeatletter

\pdfpageheight\paperheight
\pdfpagewidth\paperwidth

\usepackage[T1]{fontenc}
\usepackage[latin9]{inputenc}
\usepackage[a4paper]{geometry}
\usepackage[active]{srcltx}
\usepackage{amsmath}
\usepackage{amssymb}
\usepackage{esint}

\usepackage{xcolor}

\makeatletter

\usepackage{textcomp}

\usepackage{jheppub}

\makeatletter
\def\NAT@cmprs{\z@}
\makeatother

\usepackage{etoolbox}

\patchcmd{\maketitle}{\@fpheader}{}{}{}

\newcommand*\xbar[1]{%
	\hbox{%
		\vbox{%
			\hrule height 0.5pt 
			\kern0.3ex
			\hbox{%
				\kern-0.0em
				\ensuremath{#1}%
				\kern-0.0em
			}%
		}%
	}%
}

\usepackage{amsfonts}

\usepackage{bbm}

\newcommand{\be}{\begin{equation}}
	\newcommand{\ee}{\end{equation}}
\newcommand{\bea}{\begin{eqnarray}}
	\newcommand{\eea}{\end{eqnarray}}

\newcommand{\KM}{{k_\text{\tiny KM}}}

\def\cA{\mathcal{A}}

\def\cF{\mathcal{F}}

\def\cI{\mathcal{I}}
\def\cP{\mathcal{J}}
\def\cK{\mathcal{K}}
\def\cL{\mathcal{L}}

\def\cN{\mathcal{N}}

\def\cP{\mathcal{P}}

\title{Holographic warped CFTs from General Relativity on AdS$_3$}

\author{Kristiansen Lara and}
\author{Ricardo Troncoso}

\affiliation{Centro de Estudios Cient\'{\i}ficos (CECs),  Av. Arturo Prat 514, Valdivia, Chile}
\affiliation{Facultad de Ingenier\'ia, Universidad San Sebasti\'an, sede Valdivia, General Lagos 1163, Valdivia 5110693, Chile}

\emailAdd{klara@cecs.cl}
\emailAdd{ricardo.troncoso@uss.cl}

\preprint{}

\abstract{We introduce a new set of boundary conditions for three-dimensional Einstein gravity with negative cosmological constant. The canonical generators plainly realize the semidirect sum of Virasoro and $u(1)$ Kac--Moody algebras in its standard form, without the need of additional redefinitions of the global charges. We show that requiring Euclidean BTZ and thermal AdS$_3$ to be diffeomorphic precisely induces the warped modular $S$--transformation of the chemical potentials. The corresponding warped Cardy formula then exactly reproduces the Bekenstein--Hawking entropy, where the ground state is identified as Euclidean AdS$_3$. An explicit comparison with the boundary conditions of Brown--Henneaux and Compere--Song--Strominger is also addressed.
} 

\makeatother

\begin{document}
	\maketitle 
	
	\flushbottom
	
	\newpage{}
	
	\section{Introduction}
	
	Warped conformal field theories (WCFTs) were introduced by Hofman and Strominger in \cite{Hofman:2011zj}. Under similar reasonable and mild assumptions as in \cite{Polchinski:1987dy}, in which global Poincaré and scaling symmetries are shown to enhance to the conformal algebra in 2D, they show that just requiring global invariance under translations in space and time and a chiral scaling implies that the symmetries are enhanced either to the full conformal algebra or to the semidirect sum of a Virasoro and a $u(1)$ Kac--Moody algebra. WCFTs correspond to the latter possibility. They are non--relativistic field theories, being neither Galilean nor Carrollian, since the generator of time translations behaves as a scalar instead of a vector under the $sl(2,\mathbb{R})$ subalgebra.
	
	This class of field theories becomes of interest in different contexts, particularly for the microscopic behavior of near--extremal black holes. One of the best known examples of this sort is the near--horizon description of near extremal Kerr, whose $SL(2,\mathbb{R})\times U(1)$ isometries correspond to the global ones of the warped conformal algebra~\cite{Guica:2008mu}. Thus, WCFTs have spanned a growing activity along a variety of research lines~\cite{Castro:2015uaa,Jensen:2017tnb,Song:2017czq,Song:2019txa,Afshar:2019tvp,Chen:2020juc,Aggarwal:2022xfd,Poddar:2023ljf}, including their relationship with entanglement entropy~\cite{Anninos:2013nja,Castro:2015csg,Song:2016gtd,Apolo:2018oqv,Chen:2019xpb,Apolo:2020bld,Apolo:2020qjm,Detournay:2020vrd,Chen:2022fte} and diverse holographic realizations along different setups~\cite{Detournay:2015ysa,Donnay:2015iia,Afshar:2015wjm,Detournay:2016gao,Afshar:2016wfy,Afshar:2016kjj,Castro:2017mfj,Aggarwal:2019iay,Ciambelli:2020shy,Aggarwal:2020igb,Aggarwal:2023peg,Xu:2023jex,Detournay:2024gth,Hristov:2024cjp,Afshar:2024ess,Detournay:2024lhz,Sheikh-Jabbari:2025kjd,Despontin:2026xzg}.
	
	Despite lacking full conformal invariance, this infinite--dimensional symmetry retains strong constraining power, leading in particular to warped modular covariance and a universal asymptotic density of states through the warped version of the Cardy formula~\cite{Detournay:2012pc}.
	
	The holographic realization of WCFTs is naturally formulated in three--spacetime dimensions, and it has followed three main routes\footnote{Other holographic approaches realizing the Virasoro--Kac--Moody algebra include \cite{Compere:2009zj,Blagojevic:2009ek,Henneaux:2011hv,Song:2011sr,Detournay:2012dz,Apolo:2014tua}}: notably in the context of warped AdS$_3$ black holes in topologically massive gravity~\cite{Detournay:2012pc}; the so called ``lower-spin gravity''~\cite{Hofman:2014loa}, formulated as a Chern--Simons theory of $SL(2,\mathbb{R})\times U(1)$; and General Relativity (GR) on AdS$_3$ with the boundary conditions of Compere, Song and Strominger (CSS)~\cite{Compere:2013bya}.
	
	All of these holographic setups share a positive Virasoro central charge and a negative Kac--Moody level. Remarkably, as shown by Apolo and Song~\cite{Apolo:2018eky}, the lack of unitarity displayed by descendants of negative norm is not actually fatal. Indeed, the modular bootstrap still remains viable provided the primaries are assumed to carry positive norm.
	
	Here we propose a new set of boundary conditions for pure AdS$_3$ gravity that realize the semidirect sum of Virasoro and $u(1)$ Kac--Moody algebras directly in its standard form, with fixed central charge and level, given by
	\begin{align}
		\label{eq: Central charge and KM level}
	c=\frac{3\ell}{2G}\,,\qquad \KM=-k=-\frac{\ell}{4G}\,,
	\end{align}
	without the need of additional redefinitions of the canonical generators. The phase space includes BTZ black holes, so that in the Euclidean case, global AdS$_3$ becomes the ground state. One can then show that, requiring thermal AdS$_3$ to be diffeomorphic to Euclidean BTZ, successfully reproduces the modular properties of the chemical potentials under $S$--modular transformations, yielding to the warped Cardy formula~\cite{Detournay:2012pc}, which in our case exactly reproduces the Bekenstein--Hawking entropy of BTZ black holes.
	
	Our results provide a minimal realization of warped holography in which the bulk dynamics is just described by pure GR, so that the structure of the dual WCFT$_2$ is determined by the new boundary conditions.	
	
	\section{New boundary conditions for GR on AdS$_3$}
	
	In order to propose a set of asymptotic conditions being suitable to describe a holographic WCFT$_2$ we will follow the lines of \cite{Henneaux:2013dra,Bunster:2014mua,Perez:2016vqo} in which the fall--off of the metric is given in terms of a generic choice of lapse and shift functions. A very crucial point for our purposes is further allowing the lapse and shift to depend on the dynamical fields in a precise way at the boundary that is described in what follows. 
	
	For a localized distribution of matter, we assume that the leading terms of the asymptotic form of the metric are given by
	\begin{align}
		\label{eq: Line element}
		\begin{split}
			\dd s^2&=g_{tt}\dd t^2+2g_{t\phi}\dd t\dd\phi-\frac{\ell^2\mu'\dd t\dd\rho}{\rho}\\
			&\hspace{2cm}+\frac{\ell^2\dd\rho^2}{\rho^2}+\rho^2\left[1+\frac{\ell^2}{4\rho^2}\left(\cP^2+2\cP'\right)\right]\left[1+\frac{\ell^2}{4\rho^2}\left(\cL+\cP^2\right)\right]\dd\phi^2\,,
		\end{split}
	\end{align}
	with
	\begin{subequations}
		\begin{align}
		\begin{split}
			g_{tt}&=-\rho^2\left[\mu-\frac{\mu_{\cL}\ell^2}{4\rho^2}\left(\cL+\cP^2\right)\right]\left\{\mu_{\cL}-\frac{\ell^2}{4\rho^2}\left[\mu\left(\cP^2+2\cP'\right)-2\mu''\right]\right\}+\frac{\ell^2}{4}\mu'^2\,,
		\end{split}\\
		\begin{split}
		2g_{t\phi}&=\left(\mu-\mu_{\cL}\right)\rho^2+\frac{\ell^2}{2}\left[\mu\left(\cP^2+2\cP'\right)-\mu''-\mu_{\cL}\left(\cL+\cP^2\right)\right]\\
		&\hspace{4cm}+\frac{\ell^4}{16\rho^2}\left(\cL+\cP^2\right)\left[\left(\mu-\mu_{\cL}\right)\left(\cP^2+2\cP'\right)-2\mu''\right]\,,
		\end{split}
	\end{align}
	\end{subequations}
	where $\cL=\cL(t,\phi)$ and $\cP=\cP(t,\phi)$ stand for the dynamical fields, while the corresponding ``chemical potentials'' (sources), $\mu_{\cL}$ and $\mu_\cP$ are assumed to be fixed to a constant without variation at the boundary. Here $(\cdots)'=\partial_\phi\left(\dots\right)$.
	
	The asymptotic behavior of the metric then becomes fully specified once the remaining function $\mu$ is precisely defined. In order to carry out this task, as well as to further study the asymptotic structure, it is useful to introduce $\cI_{\cP,\cK}$ as the periodic solution of
	\begin{align}
		\label{eq: Differential equation cI}
		\left(\partial_\phi-\cP\right)\cI_{\cP,\cK}=\frac12\cK\,,
	\end{align} 
	given by\footnote{Periodicity amounts to discard the homogeneous solution of \eqref{eq: Differential equation cI}.}
	\begin{align}
		\label{eq: Integral definition cI}
		\cI_{\cP,\cK}\left(t,\phi\right):=\frac12 e^{\cF\left(t,\phi\right)}\int^\phi\dd\varphi\,e^{-\cF\left(t,\varphi\right)}\cK\left(t,\varphi\right)\,,\quad \cF'=\cP\,.
	\end{align}
	For the subsequent analysis, the relevant property of $\cI_{\cP,\cK}$ is that it satisfies the differential equation \eqref{eq: Differential equation cI}; while its explicit integral representation in \eqref{eq: Integral definition cI} is not required. 
	
	Thus, our boundary conditions become completely specified through the choice
	\begin{align}
		\label{eq: mu definition}
		\mu=-\mu_\cL-\mu_\cP\cI_{\cP,1}\,.
	\end{align}	
	The field equations in vacuum then imply that the subleading terms of the metric vanish, so that the family of spacetimes in \eqref{eq: Line element} exactly solves the Einstein equations, provided the dynamical fields are chiral
	\begin{align}
		\label{eq: Equations of motion final}
		\left(\partial_t+\mu_\cL\partial_\phi\right)\cL=\left(\partial_t+\mu_\cL\partial_\phi\right)\cP=0\,.
	\end{align}
	Note that $\cI_{\cP,\cK}$ carries this chirality when $\cK$ also does, i.e., $\left(\partial_t+\mu_\cL\partial_\phi\right)\cI_{\cP,\cK}=0$. This holds in particular for $\cI_{\cP,1}$, so that $\mu$ in \eqref{eq: mu definition} is also chiral. 
	
	
	Therefore, the boundary metric defined through $\dd s^2_{\rho\to\infty}=\rho^2\dd s_\text{b}^2$, which reads
	\begin{align}
		\label{eq: Boundary metric}
		\dd s^2_\text{b}=\left(\dd\phi+\mu\dd t\right)\left(\dd\phi-\mu_\cL\dd t\right)\,,
	\end{align}
	becomes locally flat.
	
	The BTZ black hole is recovered in the case of $\cL$ and $\cP$ constants\footnote{Note that $\cP$ can be chosen to be nonnegative $\left(\cP\geq0\right)$, since the negative branch describes the same configurations but in a rotating frame at infinity.}, provided that the following ``geometric bounds''
	\begin{align}
		\label{eq: Geometric bounds}
		\cL+\cP^2\geq0\,,\quad \cP^2\geq0\,,
	\end{align}
	are fulfilled, which saturate at the extremal cases. 
	
	We should stress that global AdS$_3$ is excluded from the boundary conditions in Lorentzian signature. Indeed, in the case of $\cL_\text{AdS}=0$ and $\cP_\text{AdS}=i$, the metric \eqref{eq: Line element} becomes complex. Nevertheless, this configuration is allowed in the Euclidean case, since the metric becomes real and fits within our asymptotic conditions (see Fig.~\ref{fig: Phase space diagram L and J2}).
	
	\begin{center}
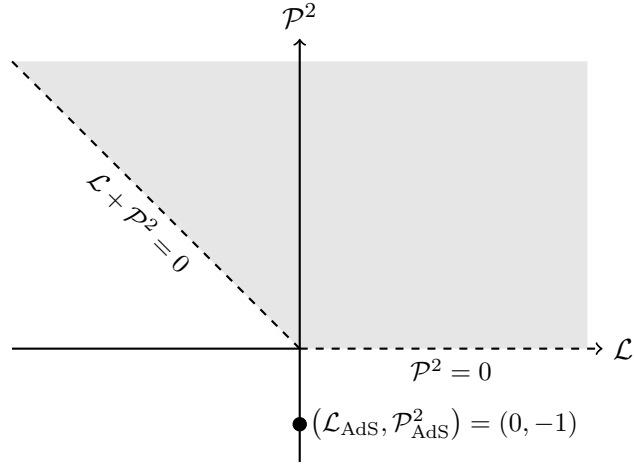

		\begin{tikzpicture}[scale=1]
			
			\begin{scope}
				\clip (-3.8,0) rectangle (3.8,3.8);
				\fill[gray!20]
				(0,0) -- (3.8,0) -- (3.8,3.8) -- (-3.8,3.8) -- (0,0) -- cycle;
			\end{scope}
			
			\draw[thick] (-3.8,0) -- (0,0);
			\draw[dashed,thick,->] (0,0)--(4,0) node[right] {$\mathcal{L}$};;
			\draw[thick,->] (0,-1.5) -- (0,4.1) node[above] {$\cP^2$};
			
			\draw[thick,dashed]
			(-3.8,3.8) -- (0,0)
			node[midway, sloped, below] {\small$\mathcal{L}+\cP^{2}=0$};
			
			\path (2,0) node[below] {\small$\cP^2=0$};
			
			\draw[fill=black] (0,-1) circle(2.5pt) node[right] {\small$\left(\cL_\text{AdS},\cP^2_\text{AdS}\right)=\left(0,-1\right)$};
			
		\end{tikzpicture}
		\captionof{figure}{The BTZ black hole is defined in the grey region, satisfying the geometric bounds $\mathcal L+\cP^2\geq0$ and $\cP^2\geq0$, whose dashed boundaries correspond to the extremal cases. The black dot denotes Euclidean $\mathrm{AdS}_3$.}
		\label{fig: Phase space diagram L and J2}
	\end{center}
	
	\subsection{Asymptotic symmetries and canonical generators of the WCFT algebra}
	
	The fall--off of the metric in \eqref{eq: Line element} is preserved by asymptotic Killing vectors $\xi=\xi^\mu\partial_\mu$ that fulfill $\delta_\xi g_{\mu\nu}=-\pounds_\xi g_{\mu\nu}$, whose leading terms are given by
	\begin{align}
		\begin{split}
			\xi^t&=\frac{\epsilon-\zeta}{\mu+\mu_\cL}+\left[\frac{\left(\mu+\mu_\cL\right)\left(\epsilon-\zeta\right)''-\left(\epsilon-\zeta\right)\mu''}{2\left(\mu+\mu_\cL\right)^2}\right]\frac{\ell^2}{\rho^2}\,,\\
			\xi^\rho&=\left[\epsilon'+\zeta'+\frac{\left(\epsilon-\zeta\right)\mu'}{\mu+\mu_\cL}\right]\frac{\rho}{2}+\frac{\ell^2\mu'\left[\left(\mu+\mu_\cL\right)\left(\epsilon-\zeta\right)''-\mu''\left(\epsilon-\zeta\right)\right]}{4\left(\mu+\mu_\cL\right)^2\rho}\,,\\
			\xi^\phi&=-\frac{\mu\epsilon+\mu_\cL\zeta}{\mu+\mu_\cL}+\frac{\ell^2\left[\left(\mu+\mu_\cL\right)\left(\mu\zeta''+\mu_\cL\epsilon''\right)+\left(\epsilon-\zeta\right)\mu\mu''\right]}{2\left(\mu+\mu_\cL\right)^2\rho^2}\,,
		\end{split}
	\end{align}
	provided that the field equations \eqref{eq: Equations of motion final} hold at infinity, and the dynamical fields transform as Virasoro--Kac--Moody currents, according to
	\begin{align}
		\delta\cL=\left(\cL'+2\cL\partial_\phi-2\partial_\phi^3\right)\epsilon-\eta'\cP\,,\quad \delta\cP=\frac12\eta'+\left(\epsilon\cP\right)'\,.
	\end{align} 
	Here, $\epsilon$ and $\eta$ are field--independent chiral parameters fulfilling,
	\begin{align}
		\label{eq: Chirality gauge parameters}
		\left(\partial_t+\mu_\cL\partial_\phi\right)\epsilon=\left(\partial_t+\mu_\cL\partial_\phi\right)\eta=0\,,
	\end{align}
	 and $\zeta$ is given by
	 \begin{align}
	 	\label{eq: Zeta}
	 	\zeta=\epsilon-\cI_{\cP,\eta+2\epsilon'}\,,
	 \end{align}
	 possessing the same chirality as in \eqref{eq: Chirality gauge parameters}.
	 
	Note that the asymptotic Killing vectors manifestly depend on the dynamical field $\cP$, which has to be carefully taken into account in order to define the canonical generators. 
	 
	 The variation of the corresponding surface integrals can be found in the Regge--Teitelboim approach~\cite{Regge:1974zd}. They are independent of the radial coordinate, and given by
	 \begin{align}
	 	\delta Q=\frac{k}{8\pi}\int \dd\phi\,\left[\epsilon\left(\delta\cL+2\cP\delta\cP\right)+2\left(\partial_\phi-\cP\right)\zeta\delta\cP\right]\,.
	 \end{align}
	 Therefore, by virtue of \eqref{eq: Zeta} and \eqref{eq: Differential equation cI}, nonlinear terms cancel out and the canonical generators integrate as
	 \begin{align}
	 	Q\left[\epsilon,\eta\right]=\frac{k}{8\pi}\int \dd\phi\,\left(\epsilon\cL-\eta\cP\right)\,.
	 \end{align}
	Expanding in modes according to $L_m=\frac{k}{8\pi}\int \dd\phi\,e^{im\phi}\cL$ and $P_m=\frac{k}{4\pi}\int \dd\phi\,e^{im\phi}\cP$,  the Poisson brackets of the conserved charges reproduce the semidirect sum of a Virasoro and a $u(1)$ Kac-Moody algebra 
	\begin{align}
		\label{eq: Algebra}
		\begin{split}
			\comm{L_m}{L_n}&=\left(m-n\right)L_{m+n}+\frac{c}{12}m^3\delta_{m+n,0}\,,\\ 	
		\comm{L_m}{P_n}&=-nP_{m+n}\,,\\
		\comm{P_m}{P_n}&=\frac{\KM}{2}m \delta_{m+n,0}\,,
		\end{split}
	\end{align}
	with $\left\{,\right\}\to i\left[,\right]$. In terms of the Newton constant and the AdS$_3$ radius, the Brown--Henneaux central charge and the Kac--Moody level are given by \eqref{eq: Central charge and KM level}. Thus, as naturally expected from different holographic WCFT$_2$ proposals~\cite{Detournay:2012pc,Compere:2013bya,Azeyanagi:2018har}, the Kac--Moody level is also found to be negative. 
	
	It is worth highlighting that the asymptotic symmetry generators plainly span the warped conformal algebra in its standard form, without requiring additional mappings of the generators.
	
	\section{Modular transformations, asymptotic growth of the number of states and black hole entropy}
	
	In the Euclidean continuation, our boundary conditions allow to recover some of the known properties of WCFTs under modular transformations in a straightforward way from purely geometric grounds in the bulk. Indeed, the ground state is clearly identified as the Euclidean AdS$_3$ geometry. Thus, following the lines of \cite{Carlip:1994gc,Maldacena:1998bw} one can prove that Euclidean AdS$_3$ and BTZ geometries with the boundary conditions in \eqref{eq: Line element} turn out to be diffeomorphic provided that the corresponding chemical potentials are related as
	\begin{align}
		\label{eq: Wmodular transformations}
		\mu_{\cL_\text{AdS}}=\frac{4\pi^2}{\mu_\cL}\,,\quad \mu_{\cP_\text{AdS}}=\frac{2i\pi\mu_\cP}{\mu_\cL}\,.
	\end{align}
	This result precisely reproduces the behavior of the temperature and angular potential under an $S$--modular transformation found in \cite{Detournay:2012pc},  that was directly obtained from a two--dimensional WCFT\footnote{Our chemical potentials relate to theirs according to $\mu_\cP=2\beta$ and $\mu_\cL=i\theta$. Note that the properties under modular transformations naturally agree with that of a chiral CFT with a $u(1)$ current, see e.g., \cite{Kraus:2006nb}.}. Thus, since the partition function of a WCFT$_2$ transforms anomalously under \eqref{eq: Wmodular transformations}, 
	\begin{align}
		Z\left(\mu_\cL,\mu_\cP\right)=e^{\frac{\KM\mu_\cP^2}{16\mu_\cL}}Z\left(\frac{4\pi^2}{\mu_\cL},\frac{2i\pi\mu_\cP}{\mu_\cL}\right)\,,
	\end{align}
	and the spectrum possesses a gap, under the standard mild assumptions one obtains that the asymptotic growth of the number of states is described by
	\begin{align}
		\label{eq: WCardy formula}
		S_\text{WCFT}=\frac{4i\pi}{\KM}P_0P_0^\text{AdS}+4\pi \sqrt{-\left(L_0-\frac{P_0^2}{\KM}\right)\left[L_0^\text{AdS}-\frac{\left(P_0^\text{AdS}\right)^2}{\KM}\right]}\,,
	\end{align}
	in full agreement with the ``warped Cardy formula'' found in \cite{Detournay:2012pc}, where in our case the ground state is precisely identified as the Euclidean AdS$_3$ geometry.
	
	For Euclidean BTZ and AdS$_3$, the zero modes are respectively given by,
	\begin{align}
		\label{eq: Zero modes}
		L_0=\frac{k}{4}\cL\,,\quad P_0=\frac{k}{2}\cP\,;\quad L_0^\text{AdS}=0\,,\quad P_0^\text{AdS}=\frac{ik}{2}\,,
	\end{align}
	so that replacing \eqref{eq: Zero modes} and the Kac--Moody level \eqref{eq: Central charge and KM level} into \eqref{eq: WCardy formula}, one obtains
	\begin{align}
		S_\text{WCFT}=k\pi\left(\cP+\sqrt{\cL+\cP^2}\right)=\frac{\cA}{4G}\,,
	\end{align}
	and hence, the Bekenstein--Hawking entropy is exactly recovered from the warped Cardy formula.
	
\section{Ending remarks}

It is worth mentioning that if one just requires that the variation of the global charges are integrable and finite, the asymptotic behavior of the metric \eqref{eq: Line element} could be slightly relaxed. However, consistency of the asymptotic structure implies that the field equations must be fulfilled near the boundary, which inevitably brings one back to the (sub)leading terms of \eqref{eq: Line element}.

A thorough analysis of the full asymptotic structure including details concerning the canonical generators, modular transformations as well as some aspects about unitarity and parity violation are going to be discussed in a forthcoming publication~\cite{Lara2027}.

Some remarks regarding the unitarity bound, given by 
\begin{align}
	\label{eq: Unitarity bound algebraic}
	L_0-\frac{P_0^2}{\KM}\geq -\frac{c}{24}\,,
\end{align}
are in order. Indeed, in the context of holographic WCFTs, as pointed out in \cite{Apolo:2018eky}, it is useful to assume that the bound still holds in the case of negative $\KM$, because in spite of the violation of unitarity, the modular bootstrap appears to work very successfully, provided that the primaries possess positive norm, which becomes compatible with states of imaginary $u(1)$ charge.

It is then worth noting that in our context, by means of \eqref{eq: Zero modes}, in the case of the black hole, the unitarity bound in \eqref{eq: Unitarity bound algebraic} reduces to
\begin{align}
	\label{eq: Unitarity bound geometric}
	\frac{c}{24}\left(\cL+\cP^2+1\right)\geq0\,,
\end{align}
which is always fulfilled by virtue of the geometric bounds in \eqref{eq: Geometric bounds}; while the unitarity bound saturates in the case of Euclidean AdS$_3$.

Besides, one may naturally wonder about the relationship between our chemical potentials $\mu_\cL$, $\mu_\cP$ and global charges $\cL$, $\cP$, with the standard ones in the canonical ensemble, i.e., with the inverse Hawking temperature $\beta$, the angular velocity of the horizon $\Omega_h$, as well as the mass $M$ and angular momentum $J$. The standard quantities naturally arise in the case of Brown--Henneaux boundary conditions~\cite{Brown:1986nw} for which the leading terms of the lapse and shift functions, respectively given by $\cN$ and $\cN^\phi$, are held fixed at infinity. The comparison with our boundary conditions can then be readily performed by extracting $\cN$ and $\cN^\phi$ from the asymptotic behavior in \eqref{eq: Line element}. For the sake of this discussion, the generic expression is not so enlightening; nonetheless, in the case of black hole thermodynamics it simplifies as
\begin{align}
	\label{eq: N and Nphi at infinity}
	\cN=\frac{\mu_\cP\ell}{4\cP}\,,\quad \cN^\phi=-\mu_\cL+\frac{\mu_\cP}{4\cP}\,,
\end{align}
where it must be emphasized that for our boundary conditions what remains fixed are the chemical potentials $\mu_\cP$ and $\mu_\cL$. The Hawking temperature and the angular velocity of the horizon are defined through $\cN=\beta$ and $\cN^\phi=\beta\Omega_h$, and therefore, the relation with our chemical potentials can be readily extracted from \eqref{eq: N and Nphi at infinity}. Consequently, changing the boundary conditions at infinity amounts to change the ensemble, so that the first law reads
\begin{align}
	\delta S=\frac k4\left(\mu_\cL\delta\cL+\mu_\cP\delta\cP\right)=\beta\left(\delta M+\Omega_h\delta J\right)\,,
\end{align}
with
\begin{align}
	M=\frac{k}{4\ell}\left(\cL+2\cP^2\right)\,,\quad J=-\frac{k}{4}\cL\,.
\end{align} 
One then obtains that the warped Cardy formula maps to the standard one, since
\begin{align}
	S_\text{WCFT}\left[L_0,P_0\right]=\frac{\cA}{4G}=S_\text{Cardy}\left[L_0^+,L_0^-\right]\,,
\end{align}
which manifestly expresses the Bekenstein--Hawking entropy in terms of the corresponding extensive variables in both ensembles, i.e., in terms of the zero modes of the Virasoro--Kac--Moody algebra in \eqref{eq: Zero modes} and those of two copies of the Virasoro algebra, given by $L_0^\pm=\frac12\left(\ell M\pm J\right)$, corresponding to the different boundary conditions.

Finally, it is also interesting to compare our boundary conditions with those proposed by CSS in \cite{Compere:2013bya}. Their variables are given by chiral functions of $x^+=\frac{t}{\ell}+\phi$, given by $\overline{L}$ and $\partial_+\overline{P}$, and an arbitrary constant $\Delta$ that is fixed without variation. The global charges manifestly depend on this constant, so that the Kac--Moody level is fixed in terms of the zero mode of the $u(1)$ generators according to $k_\text{\tiny CSS}=-4\Delta=-4\cP_0^\text{\tiny CSS}$, being positive for AdS$_3$ ($\Delta=-k/4$) and becomes negative for the black hole ($\Delta>0$). The Virasoro--Kac--Moody algebra \eqref{eq: Algebra} is then recovered along the lines of \cite{Detournay:2012pc} by means of a nonlocal redefinition of the modes, induced by state--dependent asymptotic Killing vectors, so that the level no longer depends on $\cP_0^\text{\tiny CSS}$. 

In order to make contact with our boundary conditions, it is necessary to perform a coordinate transformation so that the radial coordinates fit. Chiralities then match provided one of the chemical potentials is chosen as $\mu_\cL=-1$. The relationship with our chiral functions $\cL$ and $\cP$, and the remaining fixed chemical potential $\mu_\cP$ can be seen as follows.  By comparing the subleading relevant terms of both spatial metrics one readily finds that
\begin{align}
	\cL+\cP^2=\frac{4}{k}\overline{L}\,,
\end{align}
while the relation between the remaining variables is recovered by comparing the metrics at the boundary. Both boundary metrics are locally flat and conformally related by a chiral conformal factor according to
\begin{align}
	\dd s^2_{b,\text{CSS}}=\left(1+\partial_+\overline{P}\right)\dd s^2_b\,,
\end{align}
with $\dd s^2_b$ given by \eqref{eq: Boundary metric}, and 
\begin{align}
	\mu=\frac{\partial_+\overline{P}-1}{\partial_+\overline{P}+1}\,.
\end{align}
Since $\mu$ is defined through \eqref{eq: mu definition}, by virtue of \eqref{eq: Differential equation cI}, one finds that
\begin{align}
	\cP=-\left(\frac{\mu_\cP}{4}+\partial_+\ln\right)\left(1+\partial_+\overline{P}\right)\,.
\end{align}
This last equation implies that the CSS boundary conditions can be seen as a subsector of ours, provided that the zero mode of the $u(1)$ current $P_0$ is fixed without variation, and the remaining chemical potential is fixed according to
\begin{align}
	\mu_\cP=-\frac{8}{k}P_0=-8\sqrt{\frac{\cP_0^\text{\tiny CSS}}{k}}=-8\sqrt{\frac{\Delta}{k}}\,.
\end{align}
As a closing remark, following the lines of CSS~\cite{Compere:2013bya}, it is worth exploring the features of the classical WCFT that would emerge from the Hamiltonian reduction of General Relativity with our boundary conditions.

\section*{Acknowledgements}

We thank Hamed Adami, Gabriel Arenas--Henriquez, Stéphane Detournay, Oscar Fuentealba, Robinson Mancilla, Wei Song, and David Tempo for discussion and interesting remarks. This research has been partially supported by ANID FONDECYT grant N° 1250487. KL thanks the warm hospitality of Université Libre de Bruxelles, Tsinghua University and the Shanghai Institute for Mathematics and Interdisciplinary Sciences (SIMIS).

	\appendix

\bibliographystyle{JHEP}
\bibliography{Bibliographyv7}
	
\end{document}